\documentclass[12pt,a4paper]{article}

\usepackage{amsthm,amsmath,natbib,amssymb,amsfonts,bm, mathtools}
\usepackage{graphicx}
\usepackage{placeins}
\usepackage{epstopdf}
\usepackage{secdot}
\usepackage{longtable}
\usepackage{color}
\usepackage{float}
\usepackage{fancyvrb}
\usepackage{multirow}
\usepackage{etoolbox}
\usepackage{listings}

\usepackage{sectsty} 
\sectionfont{\fontsize{12}{16}\selectfont} 
\subsectionfont{\fontsize{12}{16}\selectfont} 

\newtheorem*{theorem*}{Theorem}

\theoremstyle{definition}

\theoremstyle{remark}

\DeclareMathAlphabet{\mathpzc}{OT1}{pzc}{m}{it}

\definecolor{mygreen}{rgb}{0,0.6,0}
\definecolor{mygray}{rgb}{0.5,0.5,0.5}
\definecolor{mediumseagreen}{rgb}{0.24, 0.7, 0.44}

\lstdefinestyle{MyRprograms}{
	language = R,
	basicstyle=\ttfamily\footnotesize,           
	otherkeywords={}, 
	numberstyle=\tiny\color{mygray},  
	stepnumber=1,                   
	numbersep=5pt,                  
	backgroundcolor=\color{white},      
	showspaces=false,               
	showstringspaces=false,         
	showtabs=false,                 
	frame=single,                   
	rulecolor=\color{black},        
	tabsize=2,                      
	captionpos=b,                   
	breaklines=true,                
	breakatwhitespace=false,        
	keywordstyle={},          
	commentstyle=\color{mediumseagreen},       
	stringstyle=\color{mygray},         
	escapeinside={},            
	morekeywords={byrow},              
	lineskip = {-1.5pt} 
}

\newcommand\Tstrut{\rule{0pt}{2.9ex}}         
\newcommand\Bstrut{\rule[-1.2ex]{0pt}{0pt}}   
\newcommand\TBstrut{\Tstrut\Bstrut} 
\begin{document}

\title{Computation of the expected value of a function of a chi-distributed random variable}

\author{Paul Kabaila        and
        Nishika Ranathunga  
        \\   
        \\
        Department of Mathematics and Statistics, La Trobe University, 
        \\
        Victoria 3086, Australia
}

\date{}

\maketitle

\begin{abstract}
\noindent We consider the problem of numerically evaluating the expected value of a 
smooth bounded function of a chi-distributed random variable, divided by the square root of the number of degrees of freedom.
This problem arises
in the contexts of simultaneous inference, the selection and ranking of populations and in the evaluation of multivariate t probabilities. It also arises
in the assessment of the coverage probability and expected volume properties of the some non-standard confidence regions. We use a  transformation put forward by Mori, followed by the application of the trapezoidal rule. This rule has the remarkable property that, for suitable integrands, it is exponentially convergent. We use it to create a nested sequence of quadrature rules, for the estimation of the approximation error, so that previous evaluations of the integrand are not wasted. The application of the trapezoidal rule requires the approximation of an infinite sum by a finite sum. We provide a new easily computed upper bound on the error of this approximation. 
Our overall conclusion is that this method is a very suitable candidate for the computation of 
the coverage and expected volume properties of non-standard confidence regions.  

\medskip

\noindent \textbf{Keywords:} Confidence interval; Coverage probability; Mixed rule transformation; Numerical integration; Trapezoidal rule

\end{abstract}

\section{Introduction}
\label{intro}
Consider the problem of finding an accurate and efficient method of numerically computing an integral of the form
\begin{equation}
\label{FormofExpectedValue}
\int_0^{\infty}
a(x)
\, f_{\nu}(x) \, dx,
\end{equation}
where $a$ is a smooth bounded real-valued function, $\nu$ is a positive integer and $f_{\nu}$ is the 
probability density function (pdf) of a random variable with the same distribution as $R / \nu^{1/2}$, where $R$ has a $\chi_{\nu}$ distribution (i.e. $R^2$ has a $\chi_{\nu}^2$ distribution). Note that $\eqref{FormofExpectedValue} = E \big(a(R/{\nu}^{1/2})\big)$, which is the expected value of a smooth bounded function of $R/{\nu}^{1/2}$. We suppose that a computer program for the accurate and efficient evaluation of $a(x)$, for any given $x > 0$, 
is either already available or can be easily written.
In other words, our focus is solely on the numerical evaluation of the integral \eqref{FormofExpectedValue}.

The evaluation of an integral of this form 
occurs in the context of simultaneous statistical inference  
and the selection and ranking of populations
(\citeauthor{Miller1981}
\citeyear{Miller1981}; 
\citeauthor{HochbergTamhane1987} \citeyear{HochbergTamhane1987};
\citeauthor{GuptaPanchapakesan2002} \citeyear{GuptaPanchapakesan2002})
and in the evaluation of central and non-central (Kshirsagar definition)
multivariate t probabilities 
(\citeauthor{DunnettSobel} \citeyear{DunnettSobel}; 
\citeauthor{Dunnett} \citeyear{Dunnett}; 
\citeauthor{GenzBretz} \citeyear{GenzBretz}),
when the method of \cite{MiwaHayterKuriki2003}, briefly described in \cite{MiMiwaHothorn2009}, 
is used to compute $a(x)$.

The evaluation of an integral of the form $\eqref{FormofExpectedValue}$ also occurs
in the computation of the coverage probabilities of post-model-selection
confidence intervals, frequentist model averaged confidence intervals and other non-standard confidence regions 
(\citeauthor{FarchioneKabaila2008} \citeyear{FarchioneKabaila2008};
\citeauthor{KabailaGiri2009JSPI}, \citeyear{KabailaGiri2009JSPI};
\citeauthor{KabailaGiri2009ANZJS}, \citeyear{KabailaGiri2009ANZJS};
\citeauthor{KabailaFarchione2012} \citeyear{KabailaFarchione2012};
\citeauthor{KabailaWelshAbeysekera2016} \citeyear{KabailaWelshAbeysekera2016};
\citeauthor{KabailaWelshMainzer2017} \citeyear{KabailaWelshMainzer2017};
\citeauthor{AbeysekeraKabaila2017} \citeyear{AbeysekeraKabaila2017};
\citeauthor{KabailaGiri2013}, \citeyear{KabailaGiri2013};
\citeauthor{KabailaTissera2014},  \citeyear{KabailaTissera2014};
\citeauthor{Kabaila2018} \citeyear{Kabaila2018}).
In all of these papers, this evaluation has previously been carried out by 
first truncating the integral (the truncation error is easily bounded) and then applying an adaptive numerical integration method.

Our search for a better method for the evaluation of an integral of the form $\eqref{FormofExpectedValue}$ has led us to seek out an appropriate  
transformation of the variable of integration, followed by the application trapezoidal rule over the real line. As noted by \cite{Schwartz1969}, ``The real artistry of numerical integration
lies in learning to make changes of the variable'' appropriate for the problem at hand and that ``this must be studied separately for every problem''. The literature on various initial changes of the variable of integration for the purpose of efficient numerical integration is very large, with early references including \cite{DavisRabinowitz1984},  \cite{SagSzekeres1964} and \cite{Imhof1963}. Some simple illustrations of the power of appropriate changes variable of integration prior to numerical integration are provided by \cite{AverySoler1988}.

We use the transformation (2.6) of \cite{Mori1988}, followed by application of the trapezoidal rule. 
This transformation belongs to a family of transformations proposed and investigated by \cite{TakahasiMori1973}, \cite{Mori1985} and others.
The trapezoidal rule has the remarkable property that, for suitable integrands, it is exponentially convergent (\citeauthor{TrefethenWeidemanSIAMReview2014}, \citeyear{TrefethenWeidemanSIAMReview2014}).
There are several well-known explanations for this 
remarkable property, including the Euler-Maclaurin summation formula and Fourier transform methods. A historical review of the these explanations is provided in Section 11 of
\cite{TrefethenWeidemanSIAMReview2014}.
The trapezoidal rule also has the great advantage that it can be used to create a nested sequence of quadrature rules, used for the estimation of the approximation error, so that previous evaluations of the function $a$ are not wasted.

For our purposes, the best description of the properties of the trapezoidal rule
is found using the Fourier transform of the integrand and the Poisson summation formula. For the reader's convenience, this well-known description is recounted in Section 2.
The application of 
the  transformation (2.6) of \cite{Mori1988}
to the integral $\eqref{FormofExpectedValue}$, followed by the application of the trapezoidal rule is described in Section 3. In this section, we describe a method of carrying out the required `trimming' of the infinite sum approximation to the integral that leads to an easily-computed upper bound on the resulting \textsl{trimming error}. In subsection 3.1, 
we describe 
a simple and effective procedure,
similar to that described by
\citeauthor{Mori1988}
(\citeyear{Mori1988}, pp.370--371), 
for evaluating the integral 
\eqref{FormofExpectedValue}
that leads to a nested sequence of quadrature rules. 
In subsection 3.2, we describe an extension of this procedure that we prove to be exponentially convergent under the appropriate regularity condition.

In Section 4 we use the simple test scenario that consists of evaluating a known univariate t probability (i.e the value of $\eqref{FormofExpectedValue}$ is known). We compare 
the performance of the method described in subection 3.1 with the following two methods:

\begin{enumerate}
	
	\item 
	
	\underline {Generalized Gauss Laguerre quadrature}
	
	Change the variable of integration 
	in \eqref{FormofExpectedValue}
	to $y =\nu \, x^2 / 2$. In effect, we express the expectation of interest, 
	$E \big(a(R/{\nu}^{1/2})\big)$, as 
	$E \big(a(2^{1/2} \, V^{1/2} / \nu^{1/2})\big)$,
	where $V = R^2/2$ has a gamma$(\nu/2, 1)$ distribution. We then apply 
	Generalized Gauss Laguerre quadrature. This method has been widely applied in the 
	literature on simultaneous inference  
	and the selection and ranking of populations.
	
	\item 
	
	\underline {Inverse cdf method}
	
	Change the variable of integration 
	in \eqref{FormofExpectedValue}
	to $y = F_{\nu}(x)$, where $F_{\nu}$ denotes the cumulative distribution function (cdf) that corresponds to the pdf $f_{\nu}$. This transforms the integral $\eqref{FormofExpectedValue}$ into an integral over the interval $[0,1]$. 
	In effect, we express the expectation of interest, 
	$E \big(a(R/{\nu}^{1/2})\big)$, as $E \big(a(F_{\nu}^{-1}(U))\big)$,
	where $U = F_{\nu}(R / \nu^{1/2})$ is uniformly distributed on $(0,1)$.
	We then apply Gauss Legendre quadrature. This method has been applied to 
	the evaluation of central and non-central (Kshirsagar definition)
	multivariate t probabilities. 
	
	%
	
\end{enumerate}

\noindent The purpose of this comparison is to illustrate the factors that may lead to a relatively poor performance of these two methods.
The computations for this paper were carried out using the \texttt{R}
computer language.

Finally, in Section 5 we discuss the application of the procedures described in Section 3 to the 
computation of the coverage probability and scaled expected length
of of post-model-selection
and frequentist model averaged confidence intervals. We also consider the application of these procedures to the computation of the coverage probability and scaled expected volume of other non-standard confidence regions.
\section{Properties of the trapezoidal rule found using the Fourier transform of the integrand}
\label{sec:PropsTrapRuleUsingFT}

Suppose that we wish to evaluate 
\begin{equation}
\label{IntegralWeWishToEvaluate}
\int_{-\infty}^{\infty} g(y) \, dy,
\end{equation}
where $g$ is a real-valued absolutely integrable function. Let $G$ denote that Fourier transform of $g$. This transform is defined by 
\begin{equation*}
G(\omega) = \int_{-\infty}^{\infty} g(y) \exp(- i \, \omega \, y) \, dy,
\end{equation*}
where $i = \sqrt{-1}$ and the angular frequency $\omega \in \mathbb{R}$. Since $g$ is real-valued,
$G(\omega)$ is an even function of $\omega$ (see e.g. p.11 of \citeauthor{Papoulis1962} \citeyear{Papoulis1962}). 
It follows from the Poisson summation formula
(see e.g. p.47 of \citeauthor{Papoulis1962} \citeyear{Papoulis1962})
that
\begin{equation}
\label{DiscretizationError}
\left|h \sum_{j = -\infty}^{\infty} g(j h + \delta)
- \int_{-\infty}^{\infty} g(y) \, dy \right|
\le 2 \sum_{j=1}^{\infty} \left | G \left(\frac{2 \pi j}{h}\right) \right |,
\end{equation}
for all $\delta \in [0, h)$.
The left-hand side is the \textsl{discretization error}. This error is small when $|G(\omega)|$ decays rapidly as $\omega \rightarrow  \infty$ and $h$ is sufficiently small.

We approximate the infinite sum
\begin{equation}
\label{InfiniteSum}
h \sum_{j = -\infty}^{\infty} g(j h + \delta)
\end{equation}
by the finite sum 
\begin{equation}
\label{FiniteSum}
h \sum_{j = M}^{N} g(j h + \delta),
\end{equation}
for appropriately chosen integers $M$ and $N$ ($M < N$). 
The ``trapezoidal rule'' approximation to \eqref{IntegralWeWishToEvaluate} is
\eqref{FiniteSum}.
The absolute value of the difference \eqref{FiniteSum} $-$ \eqref{InfiniteSum} is called the \textsl{trimming error}. 
For \eqref{FiniteSum} to be a good approximation to 
\eqref{IntegralWeWishToEvaluate}, we require that both the \textsl{discretization error}
and the \textsl{trimming error} are small.

\section{Application of the the transformation (2.6) of \cite{Mori1988},
	followed by the application of the trapezoidal rule}
\label{MoriFollowedByTrapRule}

The pdf $f_{\nu}$ is given by 
\begin{equation*}
f_{\nu}(x) = 
\begin{cases}
\tau_{\nu} \, x^{\nu - 1} \, \exp\big( - \nu \, x^2/2 \big)
&\text{for}\ \  x > 0
\\
0 &\text{otherwise,}
\end{cases}
\end{equation*}
where
\begin{equation}
\tau_{\nu} = \displaystyle{\frac{\nu^{\nu/2}}{\Gamma(\nu/2) \, 2^{(\nu/2)-1}}}. 
\end{equation}
Throughout this section we suppose that $\nu$ is given. To evaluate \eqref{FormofExpectedValue}, we first apply the transformation
(2.6) of \cite{Mori1988}, namely
\begin{equation*}
x(y) = 
\exp \left(\frac{1}{2} \, y - e^{-y }\right),
\end{equation*}
so that 
\begin{equation*}
\frac{dx(y)}{dy} = 
\exp \left(\frac{1}{2} \, y - e^{-y }\right) \, 
\left(\frac{1}{2} + e^{-y } \right)
\end{equation*}
and 
\begin{equation}
\label{MixedRuleTransfOfExpectVal}
\int_0^{\infty}
a(x)
\, f_{\nu}(x) \, dx
= \int_{-\infty}^{\infty} a\big(x(y) \big) \, \psi_{\nu}(y) \, dy,
\end{equation}
where 
\begin{equation*}
\psi_{\nu}(y) = f_{\nu} \big(x(y) \big) \frac{dx(y)}{dy}.
\end{equation*}
As noted by \cite{Mori1985} this transformation leads to
$\psi_{\nu}(y)$ having \textsl{double exponential} decay as 
$y \rightarrow \pm \infty$, i.e. there exist positive numbers $c_1$, $c_2$ and $c_3$
such that
\begin{equation}
\label{DEdecay}
|\psi_{\nu}(y)| \sim c_1 \exp\big(-c_2 \exp(c_3|y|) \big), \ y \rightarrow \pm \infty.
\end{equation}
This implies that 
$g_{\nu}(y) = a\big(x(y) \big) \, \psi_{\nu}(y)$ also has \textsl{double exponential} decay as 
$y \rightarrow \pm \infty$.
Computational results show that the function $\psi_{\nu}$ is unimodal for all positive integers $\nu$. Let $y_{\nu}^*$ denote the value of $y$ at which $\psi_{\nu}(y)$ is maximized.
The value of $y_{\nu}^*$ is roughly 0.85 for all positive integers $\nu$.
We suppose, without loss of generality, that $|a(x)| \le 1$ for all $x \in \mathbb{R}$.

Let $G_{\nu}$ denote the Fourier transform of $g_{\nu}(y)$. We now introduce the following assumption:

\smallskip

\noindent \textbf{Assumption FT:} \ There exist positive numbers $c_4$ and $c_{FT}$
such that 
\begin{equation*}
|G_{\nu}(\omega)| \le c_4 \exp\big(- c_{FT} |\omega|\big)
\end{equation*}
for all $\omega \in \mathbb{R}$. In other words, $G_{\nu}(\omega)$ has \textsl{single exponential} decay as $\omega \rightarrow \pm \infty$.

\smallskip

\noindent Theorem 5.1 of \cite{TrefethenWeidemanSIAMReview2014}
provides conditions on the function $g_{\nu}(y)$ that imply that this assumption holds. 

We will approximate \eqref{MixedRuleTransfOfExpectVal} by
\begin{equation}
\label{TrapRuleApprox}
h \, \sum_{j = 0}^{n-1} a \big( x (y_{\ell} + h j) \big) \, \psi_{\nu}(y_{\ell} + h j),
\end{equation}
where $n$ denotes the number of evaluations
of the integrand $a\big(x(y) \big) \psi_{\nu}(y)$, $h$ denotes the step length and the first evaluation of this integrand is at $y_{\ell}$. 
Let $d = n  h$. Of course, our aim is to choose 
$\big(n, h, y_{\ell}\big)$ such that \eqref{TrapRuleApprox} provides an efficient and accurate approximation. 

We will use the following result which provides an easily computed
upper bound on the \textsl{trimming error}.

\noindent \textbf{Lemma 1.} \textit{
	Suppose that $y_{\ell} < y_{\nu}^*$ and that $y_{\ell} + d > y_{\nu}^*$. Then, when we approximate  \eqref{MixedRuleTransfOfExpectVal} by \eqref{TrapRuleApprox},
	the trimming error is bounded above by $u_{\nu}\big(y_{\ell}, d \big)$,
	where
	\begin{equation*}
	u_{\nu}(y, d)
	= Q_{\nu} \Big(\nu \, x^2(y) \Big) + 1 - Q_{\nu}  \Big(\nu \, x^2 \big(y + d \big) \Big)
	\end{equation*}
	and $Q_{\nu}$ denotes the $\chi_{\nu}^2$ cdf. 
}

\noindent \textit{Proof.}
Suppose that $y_{\ell} < y_{\nu}^*$ and that $y_{\ell} + d > y_{\nu}^*$. The \textsl{trimming error} is 
\begin{equation}
\left |	h \, \sum_{j = -\infty}^{-1} a \big( x (y_{\ell} + h j) \big) \, \psi_{\nu}(y_{\ell} + h j)\, + \, h \, \sum_{j = n}^{\infty} a \big( x (y_{\ell} + h j) \big) \, \psi_{\nu}(y_{\ell} + h j) \right |
\label{TransformForTrapezRule_C}
\end{equation}
The \textsl{trimming error} is bounded above by	
\begin{equation*}
h \, \sum_{j = -\infty}^{-1} \, \psi_{\nu}(y_{\ell} + h j) \, + \, h \, \sum_{j = n}^{\infty} \, \psi_{\nu}(y_{\ell} + h j),
\end{equation*}
since, for all positive integers $\nu$, $\psi_{\nu}(y) > 0$ for 
all $y \in \mathbb{R}$.
Observe that
\begin{align*}
h \, \sum_{j = n}^{\infty} \, \psi_{\nu}(y_{\ell} + h j)
= h \, \sum_{j = 1}^{\infty} \, \psi_{\nu}(y_u + h j),
\end{align*}
where $y_u = y_{\ell} + d$.	
We now use the same reasoning as for the \textsl{integral test}
for series convergence. 
Since $\psi_{\nu}(y_u + t)$ is a decreasing function of $t \ge y_{\nu}^*$,
\begin{align*}
h \, \sum_{j = 1}^{\infty} \, \psi_{\nu}(y_u + h j) 
\leq \int_{y_u}^{\infty} \psi_{\nu}(t) \, dt 
&= \int_{y_u}^{\infty}  f_{\nu}\big(x(y)\big) \, \frac{dx(y)}{dy} \, dy
\\
&= P\big(R > \nu^{1/2} x(y_u) \big), 
\\
&= 1 - Q_{\nu}(\nu \, x^2(y_u)).
\end{align*}
Similarly, since $\psi_{\nu}(y_{\ell} + t)$ is an increasing function of $t \in (-\infty, y_{\nu}^*]$, 
\begin{equation*}
h \, \sum_{j = -\infty}^{-1} \, \psi_{\nu}(y_{\ell} + h j)
\le \int_{-\infty}^{y_{\ell}} \psi_{\nu}(t) \, dt
= Q_{\nu} \big(\nu \, x^2(y_{\ell}) \big).
\end{equation*}
Therefore \eqref{TransformForTrapezRule_C} is bounded above by
$u_{\nu}\big(y_{\ell}, d \big)$. 		
\hfill  $\square$

\subsection{A simple and effective procedure for evaluating the integral \eqref{MixedRuleTransfOfExpectVal}}

Suppose that we are given the value $\epsilon > 0$ of a desired upper bound on the absolute value of the approximation error that we will develop. We now describe 
a simple and effective procedure for evaluating the integral \eqref{MixedRuleTransfOfExpectVal}, to roughly this accuracy, 
that leads to a nested sequence of quadrature rules. This procedure,
which is similar to that described by
\citeauthor{Mori1988}
(\citeyear{Mori1988}, pp.370--371), 
consists of the following steps: 

\medskip

\noindent \underline{Step 1: Choose $y_{\ell}$ and $d$ and an initial value of $n$}

\smallskip

\noindent 
The upper bound \eqref{DiscretizationError} on the \textsl{discretization error} suggests that, for a given value of the upper bound on the 
\textsl{trimming error}, as given in Lemma 1, it makes sense to 
minimize $h$. This provides the motivation for the following choice of $d$. Choose $d$ such that 
\begin{equation*}
\min_{y} u_{\nu}\big(y, d \big) = 10^{-3} \, \epsilon.
\end{equation*}
Choose $y_{\ell}$ to be the value of $y$ minimizing $u_{\nu}\big(y, d \big)$.
This will ensure that the magnitude of the approximation error will be dominated by the \textsl{discretization error}. This is not as wasteful of evaluations of the integrand $g_{\nu}(y)$ as it might seem at first since $g_{\nu}(y)$ has \textsl{double exponential} decay as  $y \rightarrow \pm \infty$.
We have chosen the initial value of $n$ to be 5. 
Proceed to the next step.

\medskip

\noindent \underline{Step 2: For given $(n,h, y_{\ell})$, evaluate the approximation \eqref{TrapRuleApprox}}

\smallskip

\noindent Evaluate the approximation \eqref{TrapRuleApprox} and store the result. Using the stored values of the approximations decide whether or not to stop the procedure. Because the magnitude of the approximation error is dominated by the \textsl{discretization error}, this stopping rule can depend simply on estimating the \textsl{discretization errors}, as in the procedure described by
\citeauthor{Mori1988}
(\citeyear{Mori1988}, pp.370--371). Proceed to the next step.

\medskip

\noindent \underline{Step 3: Halve $h$ and go back to the previous step}

\subsection{An exponentially convergent procedure for evaluating the integral \eqref{MixedRuleTransfOfExpectVal}}

While the procedure described in the previous subsection is simple to program and effective (as evidenced by the numerical results presented in Section \ref{Comparison}), it does not lead to exponential convergence. We now describe a procedure  
that results in a nested sequence of quadrature rules
that, under Assumption FT, is exponentially convergent. 
The fact that $g_{\nu}(y)$ has \textsl{double exponential} decay as $y \rightarrow \pm \infty$, whereas its Fourier transform 
$G_{\nu}(\omega)$ has only \textsl{single exponential} decay as $\omega \rightarrow \pm \infty$, implies that, at each iteration, $d$ should be increased at a slower rate than $1/h$. By 
adding a given positive number $2 b$ to $d$ and halving $h$ at each iteration, 
we obtain exponential convergence.
For simplicity of exposition, we have not included a stopping rule in the description of this procedure.

\medskip

\noindent \underline{Step 1: An initial choice of a reasonable value of $(y_{\ell}, n, d)$}

\smallskip

\noindent Choose an initial value of $n$, which we denote by $n_0$.   The initial value of $h$, denoted by $h_0$, is the initial value of $d$ (to be specified shortly) divided by $n_0$. 
We choose $b$ to be some small positive integer multiple of $h_0$. 
For the sake of concretness, we have chosen $b = h_0$.
The initial value of $d$ is such that 
\begin{equation*}
\min_{y} u_{\nu}\left(y, d  \right) \ \text{is equal to some specified small positive number}.
\end{equation*}
The initial value of $y_{\ell}$, 
denoted by $y_{\ell0}$, is the value of $y$ minimizing 
$u_{\nu}(y, d)$ for the chosen initial value of $d$,
denoted by $d_0$. Let $y_{u0} = y_{\ell0} + d_0$.
Proceed to the next step.

\medskip

\noindent \underline{Step 2: For given $(y_{\ell}, n, d)$, evaluate the approximation \eqref{TrapRuleApprox}}

\smallskip

\noindent Evaluate the approximation \eqref{TrapRuleApprox} and store the result. Proceed to the next step.

\medskip

\noindent \underline{Step 3: Add $2 b$ to $d$, halve $h$ and choose the new value of $y_{\ell}$}

\smallskip

\noindent Add $2 b$ to $d$ and halve $h$. 
Choose the new value of $y_{\ell}$ to be
$y_{\ell} - b$. It will be convenient for the proof of exponential 
convergence to define the iteration number $k$ by $h = h_0 / 2^k$. 
Go back to the previous step.

\medskip

The following theorem states that under Assumption FT this procedure is exponentially convergent. The type of convergence described in this theorem is consistent with that other \textsl{double exponential} types of quadrature formulas (\citeauthor{MoriSugihara2001}, \citeyear{MoriSugihara2001}). 

\noindent \textbf{Theorem 1.} \textit{
	Suppose that Assumption FT holds.
	Then the magnitude of the approximation error is, for all sufficiently large iteration numbers $k$, bounded above by 
	\begin{align*}
	&\frac{10 \, \tau_{\nu}}{9 \nu} \left(
	\exp \left(- \frac{\nu}{2} \exp\left(\frac{9 \, y_{u0}}{10}\right)2^{c_T k}\right)
	+ \exp \left(- \nu \exp\left(-\frac{9 \, y_{\ell 0}}{10}\right)2^{c_T k}
	\right) \right)
	\\
	&+ 2 \, c_4 \exp \left(- \left(\frac{2 \, \pi \, c_{FT}}{h_0}\right) 2^k\right),
	\end{align*}
	where $c_T = 9 \, h_0 \big/ \big(10 \, \log_e(2)\big)$.
	Since, at iteration number $k$, $n =  (n_0 + 2 \, k ) \, 2^k$,
	the magnitude of the approximation error converges exponentially to 0
	as $n \rightarrow \infty$.
}

\noindent \textit{Proof.}
Suppose that Assumption FT holds. 
By the proof of Lemma 1, the 	\textsl{trimming error} for iteration number $k$, is bounded above by 
\begin{align}
\label{UpperBndTrimmingError}
\int_{y_{u 0}+k h_0}^{\infty} \psi_{\nu}(t) \, dt + \int_{-\infty}^{y_{\ell 0} - k h_0} \psi_{\nu}(t) \, dt.
\end{align}
It may be shown that there exist $t_1 < \infty$ and 
$t_2 > -\infty$ such that 
\begin{equation*}
\psi_{\nu}(t) \le \frac{\tau_{\nu}}{2} \exp \left( - \frac{\nu}{2} \exp\left(\frac{9}{10} t \right)\right)
\quad \text{for all} \quad t \ge t_1
\end{equation*}
and
\begin{equation*}
\psi_{\nu}(t) \le \tau_{\nu} \exp \left( - \nu \exp\left(-\frac{9}{10} t \right)\right)
\quad \text{for all} \quad t \le t_2.
\end{equation*}
It follows from this that
\begin{equation*}
\int_y^{\infty} \psi_{\nu}(t) \, dt 
\le \frac{10 \, \tau_{\nu}}{9 \, \nu}
\exp \left(-\frac{\nu}{2} \exp \left(\frac{9}{10} y\right)\right)
\quad \text{for all} \quad y \ge t_1
\end{equation*}
and 
\begin{equation*}
\int_{-\infty}^y \psi_{\nu}(t) \, dt 
\le \frac{10 \, \tau_{\nu}}{9 \, \nu}
\exp \left(- \nu \exp \left(-\frac{9}{10} y\right)\right)
\quad \text{for all} \quad y \le t_2.
\end{equation*}
Therefore, for all sufficiently large iteration numbers $k$, \eqref{UpperBndTrimmingError}
is bounded above by
\begin{align*}
\frac{10 \, \tau_{\nu}}{9 \nu} \left(
\exp \left(- \frac{\nu}{2} \exp\left(\frac{9 \, y_{u0}}{10}\right)2^{c_T k}\right)
+ \exp \left(- \nu \exp\left(-\frac{9 \, y_{\ell 0}}{10}\right)2^{c_T k}
\right) \right),
\end{align*}
where $c_T = 9 h_0 \big/ \big(10 \, \log_e(2)\big)$.

It follows from the upper bound \eqref{DiscretizationError} on the \textsl{discretization error} 
and
Assumption FT that, for all sufficiently large iteration numbers $k$, the
\textsl{discretization error} is bounded above by 
\begin{equation*}
2 \, c_4 \exp \left(- \left(\frac{2 \, \pi \, c_{FT}}{h_0}\right) 2^k\right).
\end{equation*}
\hfill $\square$

\section{Comparison with two other methods of numerical integration}
\label{Comparison}

In this section we use the simple test scenario that consists of evaluating a known univariate t probability. We compare 
the performance of the method described in the previous section with the two other methods described in the introduction.

\begin{figure}[H]
	\centering
	\includegraphics[scale=0.55]{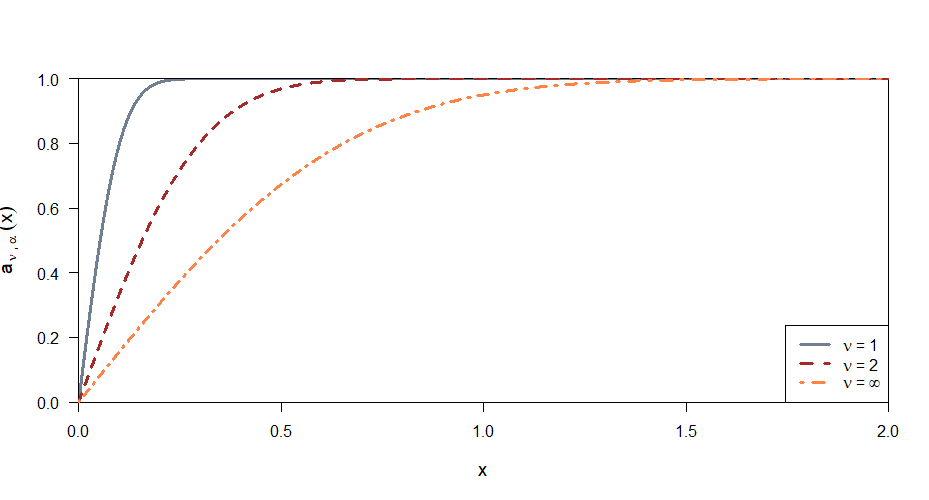}
	\caption{Graphs of $a_{\nu, \alpha} (x) $ as a function of $x$ for 
		$\alpha =0.05$ and 
		$\nu = 1, 2$ and $\infty$}
	\label{GraphsOfFunction_a}
\end{figure}

Through the consideration of the coverage probability of a $1 - \alpha$ 
t-interval, it may be shown that
\begin{align}
\label{KnownCoverageSimpler}
1-\alpha
= \int_0^{\infty}
a_{\nu, \alpha}(x)
\, f_{\nu}(x) \, dx,
\end{align}
where
\begin{equation*}
a_{\nu, \alpha}(x) = 2 \, \Phi(t_{\nu, 1-\alpha/2} \, x) - 1,
\end{equation*}
with $\Phi$ the $N(0,1)$ cdf and the quantile $t_{\nu, a}$ defined by 
$P \big(T \le t_{\nu, a}\big) = a$ for $T \sim t_{\nu}$.
Figure \ref{GraphsOfFunction_a} provides an illustration of the fact that the 
$a_{\nu, \alpha}(x)$'s are smooth bounded functions of $x$ for the values of $\alpha$ considered and all positive integers $\nu$. This figure presents
graphs of $a_{\nu, \alpha} (x) $ as a function of $x$ for 
$\alpha =0.05$ and 
$\nu = 1, 2$ and $\infty$. The graph labeled $\nu = \infty$ refers to the case that
$t_{\nu, 1-\alpha/2}$ is replaced by its limit, as $\nu \rightarrow \infty$.

\subsection{The transformation
	(2.6) of \cite{Mori1988}, followed by the application of the trapezoidal rule}

Apply the transformation
(2.6) of \cite{Mori1988}, so that
\begin{equation}
\label{MixedRuleTransfOfExpectValExample}
\int_0^{\infty}
a_{\nu, \alpha}(x)
\, f_{\nu}(x) \, dx
= \int_{-\infty}^{\infty} g_{\nu, \alpha} (y) \, dy,
\end{equation}
where $g_{\nu, \alpha} (y) = a_{\nu, \alpha}\big(x(y) \big) \, \psi_{\nu}(y)$.

%
%

We apply the simple procedure described in Section 
\ref{MoriFollowedByTrapRule}, with 
$\epsilon = 10^{-17}$ and stopped after the computation of the approximation for $n = 65$ for $\nu = 1$ and $n = 33$ for $\nu = 2, 3, 4, 5, 10, 100$ and 1000. 
The approximation error is defined to be this approximation minus $1 - \alpha$. Table \ref{table:TrapezoidError} presents the approximation error for $\alpha = 0.10, 0.05$ and 0.02 and $\nu = 1, 2, 3, 4, 5, 10, 100$ and 1000. 
Due to the finite precision of our computations in \texttt{R}, we interpret an entry 0 in this table as $|\text{approximation error}| < 1.11 \times 10^{-16}$.

%
\begin{table}[H] 
	\caption{The approximation error for the simple procedure described in Section 
		\ref{MoriFollowedByTrapRule}, with 
		$\epsilon = 10^{-17}$ and stopped after the computation of the approximation for $n = 65$ for $\nu = 1$ and $n = 33$ for $\nu = 2, 3, 4, 5, 10, 100$ and 1000. Here 
		$\alpha = 0.10, 0.05$ and 0.02 and $\nu = 1, 2, 3, 4, 5, 10, 100$ and 1000. We interpret an entry 0 in this table as $|\text{approximation error}| < 1.11 \times 10^{-16}$.} 
	\label{table:TrapezoidError} 
	\begin{tabular}{@{\extracolsep{0pt}} ccccccccc} 
		\hline 
		& $\nu=1$ & $\nu=2$ & $\nu=3$ & $\nu=4$ & $\nu=5$ \\ 
		\hline
		\TBstrut
		$\alpha=0.10$ & $-1.11 \times 10^{-16}$ & $-2.22 \times 10^{-16}$ & $0$ & $-2.22 \times 10^{-16}$ & $-1.11 \times 10^{-16}$ \\ 
		$\alpha=0.05$ & $9.66 \times 10^{-15}$ & $2.10 \times 10^{-14}$ & $-1.11 \times 10^{-16}$ & $-1.11 \times 10^{-16}$ & $-1.11 \times 10^{-16}$ \\ 
		$\alpha=0.02$ & $-1.23 \times 10^{-12}$ & $5.82 \times 10^{-11}$ & $9.99 \times 10^{-16}$ & $-1.11 \times 10^{-16}$ & $0$ \\ 
		\hline 
	\end{tabular}
	\begin{tabular}{@{\extracolsep{0.1pt}} ccccccccc} 
		\hline 
		& $\nu=10$ & $\nu=100$ & $\nu=1000$ \\ 
		\hline 
		\TBstrut
		$\alpha=0.10$ & $2.00 \times 10^{-15}$ & $2.78 \times 10^{-14}$ & $-2.37 \times 10^{-13}$ \\ 
		$\alpha=0.05$ & $2.11 \times 10^{-15}$ & $2.94 \times 10^{-14}$ & $-2.49 \times 10^{-13}$ \\ 
		$\alpha=0.02$ & $2.22 \times 10^{-15}$ & $3.03 \times 10^{-14}$ & $-2.57 \times 10^{-13}$ \\ 
		\hline 
	\end{tabular}  
\end{table}

\subsection{Generalized Gauss Laguerre quadrature}
\label{GeneralizedGaussLaguerreQuadrature}

To apply Generalized Gauss Laguerre quadrature to the evaluation of \eqref{FormofExpectedValue}, change the variable of integration to $y = \nu \, x^2 /2$, so that
\begin{equation*}
\int_0^{\infty}
a(x)
\, f_{\nu}(x) \, dx
=
\frac{1}{\Gamma (\nu/2)} \, \int_0^{\infty}
d_{\nu}(y)
\, c(y) \, dy,
\end{equation*}
where $c(y) = y^{(\nu/2) - 1} \, \exp(-y)$
and $d_{\nu}(y) = a\big( (2 y /\nu)^{1/2} \big)$.
We then apply Generalized Gauss Laguerre quadrature, with $m$ nodes (samples),
to approximate
\begin{equation*}
\int_0^{\infty}
d_{\nu}(y)
\, c(y) \, dy 
\end{equation*}
by
\begin{equation}
\label{GenGaussLaguerreApprox}
\sum_{j=1}^m w_j \, d_{\nu}(y_j)
\end{equation}
for the appropriately chosen $w_j$'s (which are all positive) and $y_j$'s 
($0 < y_1 < \dots < y_m < \infty$). 
We define the approximation error to be \eqref{GenGaussLaguerreApprox} minus $1 - \alpha$. 

\begin{figure}[t]
	\centering
	\includegraphics[scale=0.6]{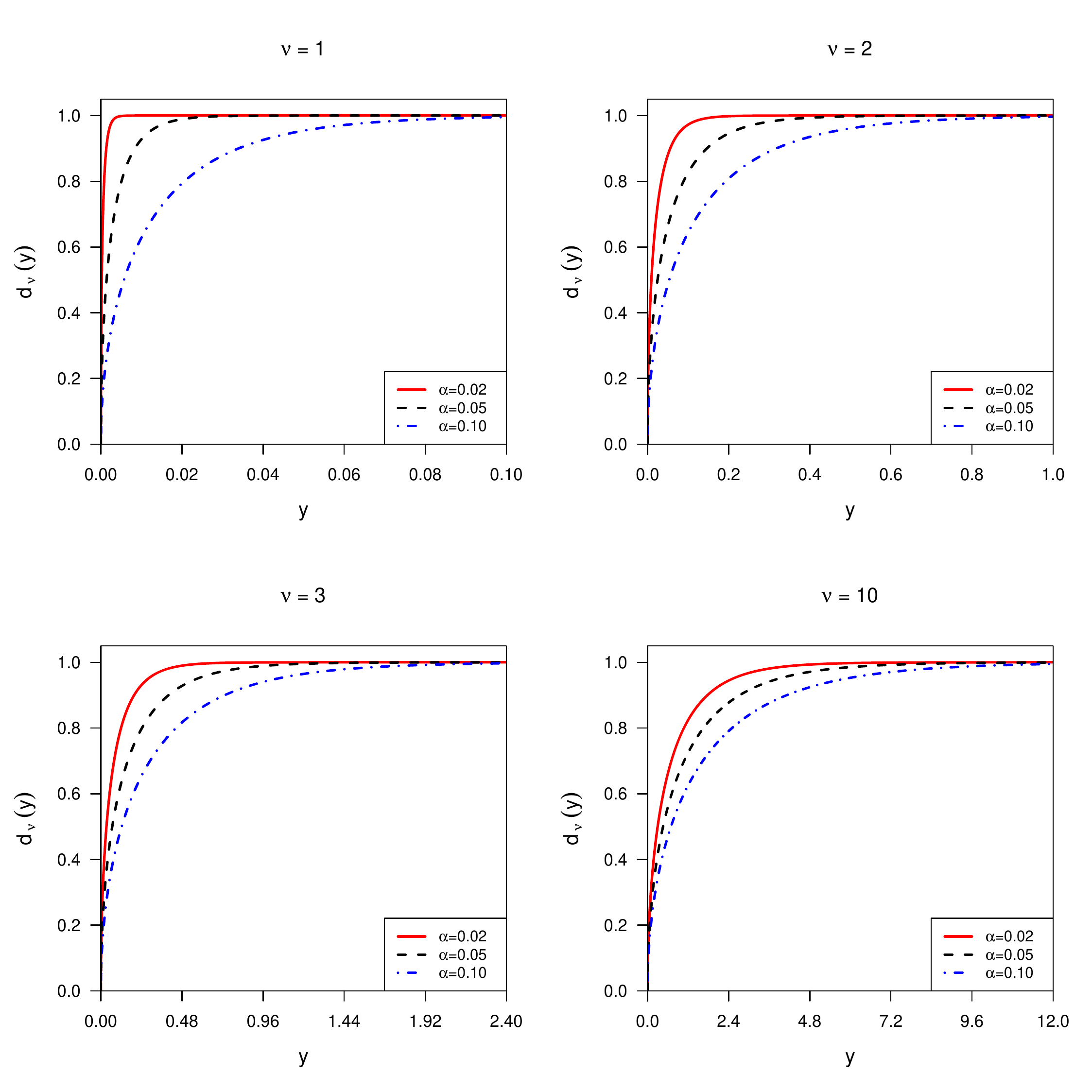}
	\caption{Graphs of $d_{\nu}(y) $ as a function of $y$ for $\nu = 1, 2, 3$ and 10 and $\alpha = 0.10, 0.05$ and 0.02}
	\label{GenGaussLaguerreGraphsOfFunction_d}
\end{figure}

Graphs of $d_{\nu}(y)$ as a function of $y$ are shown in Figure  	\ref{GenGaussLaguerreGraphsOfFunction_d}
for $\nu = 1, 2, 3$ and 10 and $\alpha = 0.10, 0.05$ and 0.02. It should be noted
that the horizontal scales in each of the four panels of this figure are very different. It is known that Generalized Gauss Laguerre quadrature with $m$ nodes will lead to the exact result if $d_{\nu}(y)$ is a polynomial in $y \in [0, \infty)$ of degree
$2m - 1$ (\citeauthor{Chandrasekhar1960},
\citeyear{Chandrasekhar1960}, p.65).

\FloatBarrier

\FloatBarrier

We assess how well $d_{\nu}(y)$ can be approximated by a polynomial, over the finite interval of values of $y$ such that $c(y) / \Gamma(\nu/2)$ is substantially greater than 0, as follows. Any polynomial $p$ 
of degree $u$ 
can be written as
\begin{equation*}
p(y) = a_0 - \sum_{j=1}^u a_j (1 - y)^j.
\end{equation*}
Set $a_0 = 1$ and require that $\sum_{j=1}^u a_j = 1$, so that the functions $p$
and $d_{\nu}$ take the same values at both $y = 0$ and $y = 1$. A first approximation to $d_{\nu}(y)$ by $p(y)$ over the interval $y \in [0,1]$ is obtained by 
minimizing a measure of distance between $d_{\nu}(y)$ and $1 - (1 - y)^j$, over
$j \in \{1, \dots, u\}$. A better approximation 
is obtained by minimizing 
a measure of distance between $d_{\nu}(y)$ and $1 - \sum_{j=1}^u a_j (1 - y)^j$, over $a_1, \dots, a_u$, subject to $\sum_{j=1}^u a_j = 1$.
It follows from the shapes of the graphs in Figure 3 that to approximate $d_{\nu}(y)$
well by a polynomial, over the finite interval of values of $y$ such that $c(y) / \Gamma(\nu/2)$ is substantially greater than 0,  we would require this polynomial to be of very high degree, 
particularly for small $\nu$.
This suggests that Generalized Gauss Laguerre quadrature, with a given number of nodes $m$,  will be most inaccurate for $\nu=1$ and will have increasing accuracy as $\nu$ increases.

This suggested result is borne out by Table 	\ref{table:GaussLguerreError} , which lists the approximation error for
Generalized Gauss Laguerre quadrature for $\alpha = 0.10, 0.05$ and 0.02 and $\nu = 1, 2, 3, 4, 5, 6, 10, 100$ and 300. 
We have chosen the number of nodes $m$ to be the same as the number of integrand evaluations in Table 	\ref{table:TrapezoidError}. In other words, the number of nodes
$m$  is $65$ for $\nu = 1$ and $33$ for $\nu = 2, 3, 4, 5, 6, 10, 100$ and 300.

%
\begin{table}[h] 
	\caption{The approximation error for Generalized Gauss Laguerre quadrature for $\alpha = 0.10, 0.05$ and 0.02 and $\nu = 1, 2, 3, 4, 5, 6, 10, 100$ and 300. The number of nodes
		$m$  is $65$ for $\nu = 1$ and $33$ for $\nu = 2, 3, 4, 5, 6, 10, 100$ and 300.} 
	\label{table:GaussLguerreError} 
	\begin{tabular}{@{\extracolsep{0pt}} ccccccccc} 
		\hline 
		& $\nu=1$ & $\nu=2$ & $\nu=3$ & $\nu=4$ & $\nu=5$ \\ 
		\hline 
		$\alpha=0.10$ & $1.44 \times 10^{-2}$ & $1.32 \times 10^{-3}$ & $1.63 \times 10^{-4}$ & $2.86 \times 10^{-5}$ & $6.08 \times 10^{-6}$ \\ 
		$\alpha=0.05$ & $3.25 \times 10^{-2}$ & $2.04 \times 10^{-3}$ & $2.26 \times 10^{-4}$ & $3.77 \times 10^{-5}$ & $7.84 \times 10^{-6}$ \\ 
		$\alpha=0.02$ & $2.00 \times 10^{-2}$ & $4.12 \times 10^{-3}$ & $3.39 \times 10^{-4}$ & $5.24 \times 10^{-5}$ & $1.05 \times 10^{-5}$ \\ 
		\hline 
	\end{tabular}
	\begin{tabular}{@{\extracolsep{0.1pt}} ccccccccc} 
		\hline 
		& $\nu=6$ & $\nu=10$ & $\nu=100$ & $\nu=300$ \\ 
		\hline 
		$\alpha=0.10$ & $1.48 \times 10^{-6}$ & $1.23 \times 10^{-8}$ & $-2.00 \times 10^{-14}$ & $-6.22 \times 10^{-15}$ \\ 
		$\alpha=0.05$ & $1.88 \times 10^{-6}$ & $1.52 \times 10^{-8}$ & $-2.11 \times 10^{-14}$ & $-6.21 \times 10^{-15}$ \\ 
		$\alpha=0.02$ & $2.46 \times 10^{-6}$ & $1.91 \times 10^{-8}$ & $-2.18 \times 10^{-14}$ & $-6.43 \times 10^{-15}$ \\ 
		\hline 
	\end{tabular}  
\end{table}

\FloatBarrier

\begin{figure}[h]
	\centering
	\includegraphics[scale=0.5]{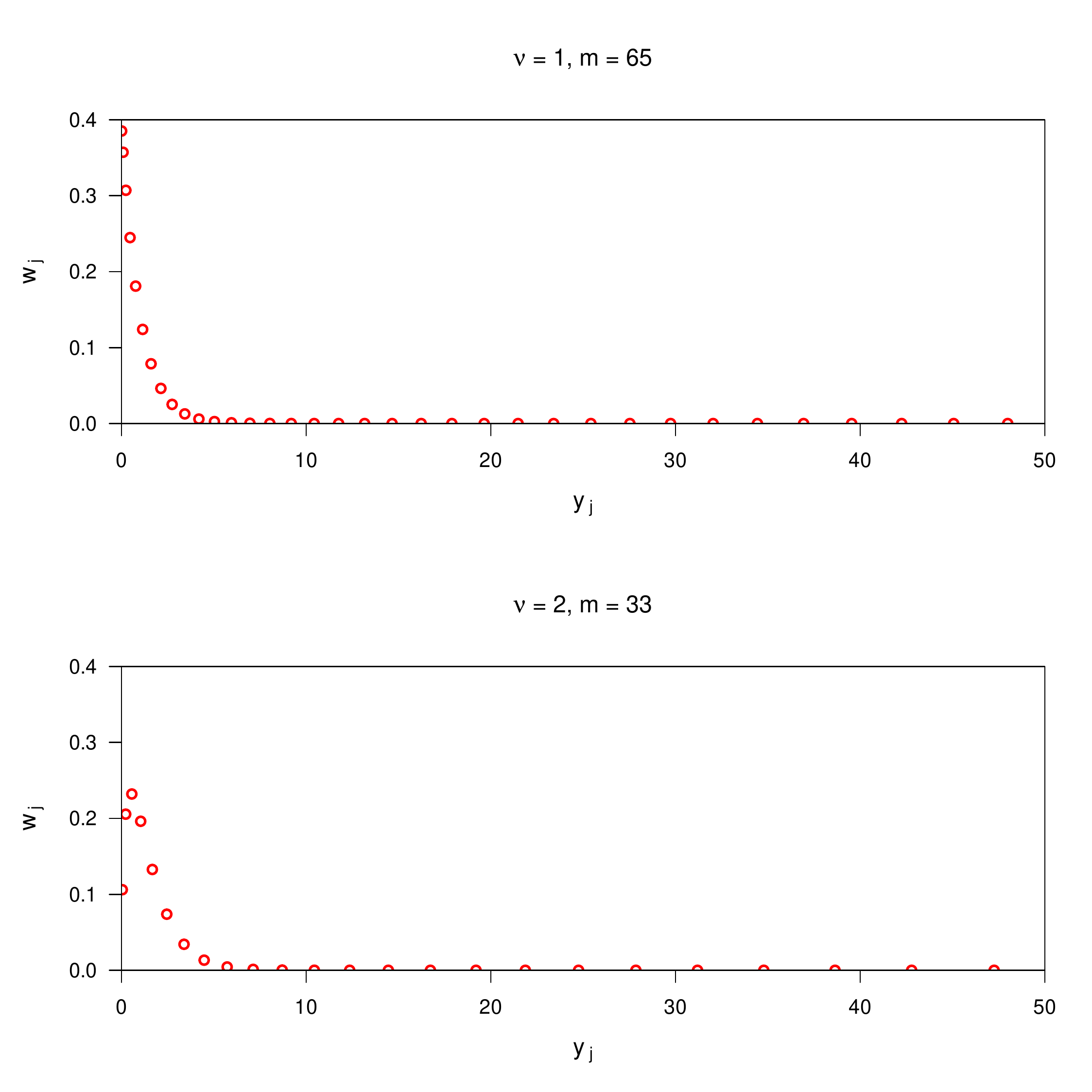}
	\caption{The top and bottom panels are scatterplots
		of the $(y_j, w_j)$'s for $(\nu, m) = (1, 65)$ and $(\nu, m) = (2, 33)$, respectively}
	\label{Scatterplots}
\end{figure}

Further confirmation of the unsuitability of Generalized Gauss Laguerre quadrature,
in the scenario under consideration, for $\nu = 1$ and $\nu = 2$ is provided by Figure 	\ref{Scatterplots}. The top and bottom panels of this figure are scatterplots
of the $(y_j, w_j)$'s for $(\nu, m) = (1, 65)$ and $(\nu, m) = (2, 33)$, respectively ($y_j \le 50$). 
For $(\nu, m) = (1, 65)$ and $(\nu, m) = (2, 33)$ there are 30 values of $y_j > 50$ and 9 values of $y_j > 50$, respectively. 
When we compare the top panel of Figure \ref{Scatterplots} with the top left panel (the case $\nu = 1$) of Figure 
\ref{GenGaussLaguerreGraphsOfFunction_d}, we
observe the following. 
Generalized Gauss Laguerre quadrature uses very few samples for the values of $y$ where the function $d_{\nu}(y)$ is changing rapidly with increasing $y$, while using a large number of samples for values of $y$ at which this function hardly changes with increasing $y$. Indeed, for $(\nu, m) = (1, 65)$  there are only 2
nodes in the interval $[0, 0.1]$. A similar conclusion results from comparing the bottom panel of Figure \ref{Scatterplots} with the top right panel (the case $\nu = 2$) of Figure 
\ref{GenGaussLaguerreGraphsOfFunction_d}. For 
$(\nu, m) = (2, 33)$  there are only 3
nodes in the interval $[0, 1]$.


\FloatBarrier

Of course, one could greatly increase the number of nodes $m$
and then approximate 
\eqref{GenGaussLaguerreApprox} by $\sum_{j=1}^q w_j \, d_{\nu}(y_j)$,
where $q$ is much less than $m$. This is unsatisfactory for the following two reasons. Firstly, the 
raison d'etre of Generalized Gauss Laguerre quadrature is that with $m$ nodes 
it leads to the exact result for polynomials of degree $2 m - 1$. This fundamental property is lost when this approximation is carried out. Secondly, this is a rather ad hoc way of forcing more samples of the 
function $d_{\nu}(y)$ into the quadrature formula for the values of $y$ for which this function changes rapidly with increasing $y$.

\subsection{Inverse cdf method, using Gauss Legendre quadrature}

Change the variable of integration to $y = F_{\nu}(x)$, where $F_{\nu}$ denotes the cdf corresponding to the pdf $f_{\nu}$,
so that
\begin{equation*}
\int_0^{\infty} a(x) \, f_{\nu}(x) \, dx
= \int_0^1 a \left(F_{\nu}^{-1}(y)\right) \, dy.
\end{equation*}
A similar transformation is used, for example, by  \citeauthor{GenzBretz} (\citeyear{GenzBretz},
p.32).
If desired, we can compute $F_{\nu}^{-1}(y)$ using either 
$F_{\nu}^{-1}(y) = \big(Q_{\nu}^{-1}(y) \big/ \nu\big)^{1/2}$
or $F_{\nu}^{-1}(y) = F_R^{-1}(y) \big / \nu^{1/2}$,
where $F_R$ denotes the $\chi_{\nu}$ cdf of $R$.
We then change the variable of integration to $z = 2 y - 1$ 
to obtain
\begin{equation*}
\int_0^1 a \left(F_{\nu}^{-1}(y)\right) \, dy
= \int_{-1}^1 b_{\nu}(z) \, dz,
\end{equation*}
where $b_{\nu}(z) = a \left(F_{\nu}^{-1}((z+1)/2)\right) \big/ 2$.
We then approximate the right-hand side, using
Gauss Legendre quadrature with $m$ nodes,
by
\begin{equation}
\label{InvCDF_GaussLegendreApprox}
\sum_{j=1}^m \widetilde{w}_j \, b_{\nu}(z_j)
\end{equation}
for the appropriately chosen $\widetilde{w}_j$'s (which are all positive) and $z_j$'s 
($-1 < z_1 < \dots < z_m < 1$). 
We define the approximation error to be \eqref{InvCDF_GaussLegendreApprox} minus $1 - \alpha$. 

\begin{figure}[h]
	\centering
	\includegraphics[scale=0.6]{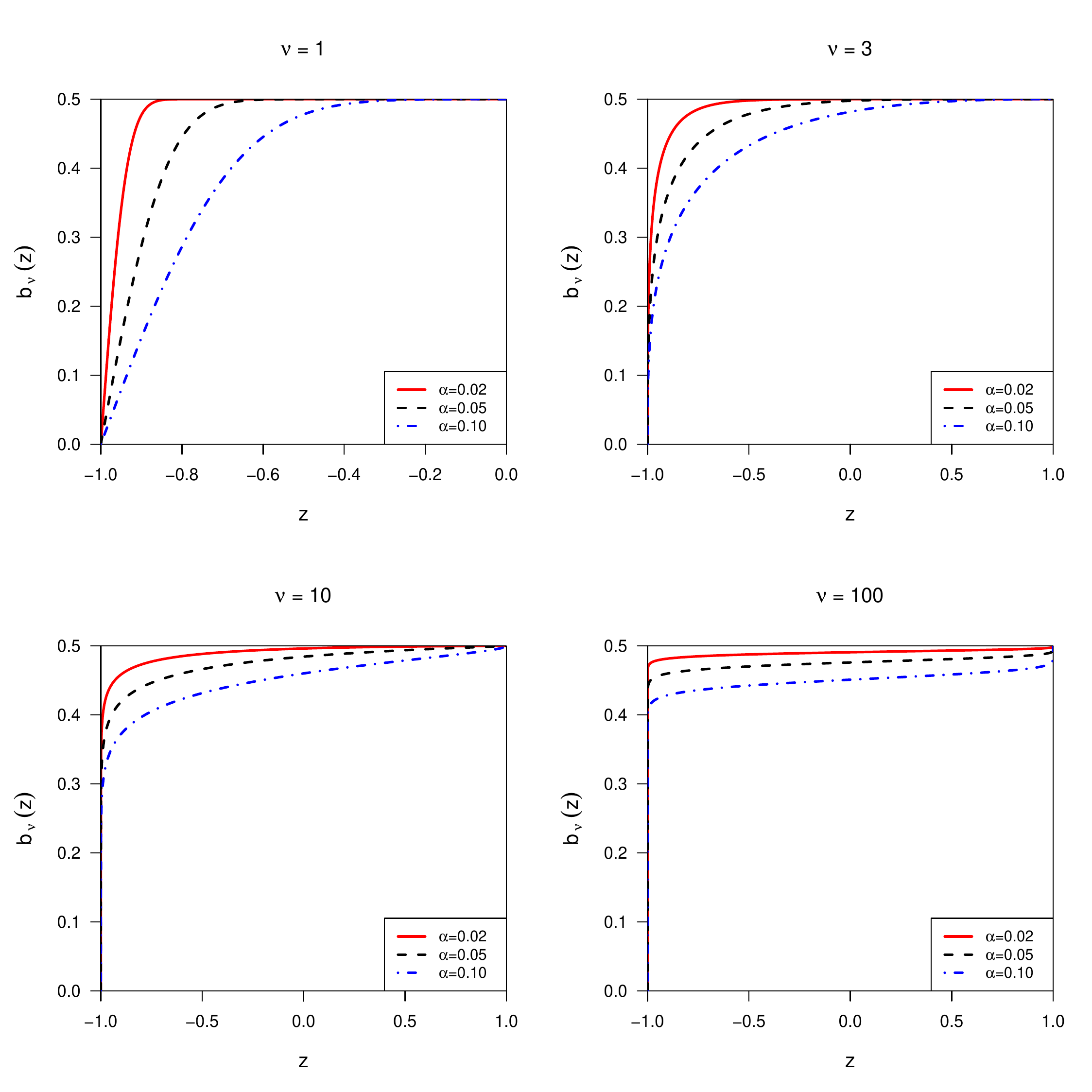}
	\caption{Graphs of $b_{\nu}(z) $ as a function of $z$ for $\nu = 1, 3, 10$ and 100 and $\alpha = 0.10, 0.05$ and 0.02}
	\label{GraphsOfFunction_b}
\end{figure}

Graphs of $b_{\nu}(z)$ as a function of $z$ are shown in Figure \ref{GraphsOfFunction_b} for 
for $\nu = 1, 3, 10$ and 100 and $\alpha = 0.10, 0.05$ and 0.02. 
It should be noted
that the horizontal scale for the $\nu = 1$ panel is different from the horizontal scale for the $\nu = 3$, $\nu = 10$ and $\nu =100$ panels (which are the same). It is known that Gauss Legendre quadrature with $m$ nodes will lead to the exact result if $b_{\nu}(z)$ is a polynomial in $z \in [-1, 1]$ of degree
$2 m - 1$.
When interpreting Figure \ref{GraphsOfFunction_b}, it is important to remember
that $b_{\nu}(-1) = 0$ and that $b_{\nu}(z)$ is an increasing continuous function
of $z \in [-1,1]$. It is evident, then, from this figure that $b_{\nu}(z)$
increases very rapidly as $z$ increases from zero for $\nu = 10$ and
$\nu = 100$.

It follows from Figure \ref{GraphsOfFunction_b}
and the same kinds of considerations as in subsection  
\ref{GeneralizedGaussLaguerreQuadrature}  
that the degree of the polynomial in $z$ needed to approximate $b_{\nu}(z)$ well in the interval 
$z \in [-1, 1]$ increases with increasing $\nu$. 
This suggests that the inverse cdf method, using Gauss Legendre quadrature with a given number of nodes $m$,  will be most accurate for $\nu=1$ and will have decreasing accuracy as $\nu$ increases.

\FloatBarrier

\FloatBarrier

This suggested result is borne out by the first 7 columns 
(the columns labelled $\nu=1$ to $\nu=10$)
of Table \ref{table:GaussLegendreError}, which lists the approximation error for
Gauss Legendre quadrature for $\alpha = 0.10, 0.05$ and 0.02 and $\nu = 1, 2, 3, 4, 5, 6, 10, 100$ and 1000. 
We have chosen the number of nodes $m$ to be the same as the number of integrand evaluations in Table \ref{table:TrapezoidError}. In other words, the number of nodes
$m$  is $65$ for $\nu = 1$ and $33$ for $\nu = 2, 3, 4, 5, 6, 10, 100$ and 1000. 

\FloatBarrier

%
\begin{table}[h] 
	\caption{The approximation error for the inverse cdf method, using Gauss Legendre quadrature, for $\alpha = 0.10, 0.05$ and 0.02 and $\nu = 1, 2, 3, 4, 5, 6, 10, 100$ and 1000. The number of nodes
		$m$  is $65$ for $\nu = 1$ and $33$ for $\nu = 2, 3, 4, 5, 6, 10, 100$ and 1000.} 
	\label{table:GaussLegendreError} 
	\begin{tabular}{@{\extracolsep{0pt}} ccccccccc} 
		\hline 
		& $\nu=1$ & $\nu=2$ & $\nu=3$ & $\nu=4$ & $\nu=5$ \\ 
		\hline 
		$\alpha=0.10$ & $7.77 \times 10^{-16}$ & $6.39 \times 10^{-6}$ &  $1.52 \times 10^{-5}$ &  $2.07 \times 10^{-5}$ &  $2.37 \times 10^{-5}$ \\ 
		$\alpha=0.05$ & $8.88 \times 10^{-16}$ & $9.43 \times 10^{-6}$ &  $2.06 \times 10^{-5}$ &  $2.70 \times 10^{-5}$ &  $2.96 \times 10^{-5}$ \\ 
		$\alpha=0.02$ & $8.88 \times 10^{-16}$ & $1.53 \times 10^{-5}$ &  $2.94 \times 10^{-5}$ &  $3.58 \times 10^{-5}$ &  $3.75 \times 10^{-5}$ \\ 
		\hline 
	\end{tabular}
	\begin{tabular}{@{\extracolsep{0.1pt}} ccccccccc} 
		\hline 
		& $\nu=6$ & $\nu=10$ & $\nu=100$ & $\nu=1000$ \\ 
		\hline 
		$\alpha=0.10$ & $2.47 \times 10^{-5}$ &  $2.30 \times 10^{-5}$ &  $4.01 \times 10^{-6}$ &  $4.23 \times 10^{-7}$ \\ 
		$\alpha=0.05$ & $3.02 \times 10^{-5}$ &  $2.64 \times 10^{-5}$ &  $4.06 \times 10^{-6}$ &  $4.23 \times 10^{-7}$ \\ 
		$\alpha=0.02$ & $3.68 \times 10^{-5}$ &  $2.90 \times 10^{-5}$ &  $3.36 \times 10^{-6}$ &  $3.33 \times 10^{-7}$ \\ 
		\hline 
	\end{tabular}  
\end{table}

\FloatBarrier

\section{Application to the computation of the coverage probability and scaled expected volume of non-standard confidence regions}

To assess the coverage probability and expected volume properties of the non-standard confidence regions considered in the references co-authored
with Kabaila, one needs to evaluate, for given $\nu$, integrals of the form \eqref{FormofExpectedValue} for hundreds, or thousands or even tens of thousands
of different functions $a$. Each of these functions is smooth and bounded and the evaluation of $a(y)$ for any given $y$ is computationally expensive.
In this case, the following ``set-up costs'' are negligible:
\begin{enumerate}
	
	\item 
	
	For the simple procedure described in Section 3 (the transformation
	(2.6) of \cite{Mori1988}, followed by application of the trapezoidal rule), the ``set-up cost'' is computing $y_{\ell}$ and $d$.
	
	\item 
	
	For Generalized Gauss Laguerre quadrature, the ``set-up cost'' is computing
	the weights, $w_j$'s, and nodes, $y_j$'s, for this quadrature, followed by the computation of the  $(2 y_j /\nu)^{1/2}$'s 
	
	\item 
	
	For the Inverse cdf method, the ``set-up cost'' consists of computing
	the weights, $w_j$'s, and nodes, $z_j$'s, for Gauss Legendre quadrature, followed by the computation of the $F_{\nu}^{-1}((z_j+1)/2)$'s.

\end{enumerate}

\noindent In other words, the number of evaluations of the function $a$ provides a reasonable guide to the computational effort for each of these methods.

We now consider in detail the evaluations of integrals of the form \eqref{FormofExpectedValue} in the references co-authored with Kabaila.
\cite{KabailaGiri2009JSPI}, \cite{KabailaGiri2009ANZJS}, 
\cite{KabailaTissera2014} and \cite{AbeysekeraKabaila2017} need to evaluate
integrals of the form 
\begin{equation*}
\int_0^{\infty} \lambda(x) \, x^{\xi} \, f_{\kappa}(x) \, dx
\end{equation*}
where $\xi$ and $\kappa$ are a positive integers and $\lambda: [0, \infty) \rightarrow \mathbb{R}$ is a smooth bounded function. This integral can be 
converted into the form \eqref{FormofExpectedValue} by changing the variable of integration to $y = c(\kappa, \xi) \, x$, where
$c(\kappa, \xi) = \big(\kappa / (\kappa + \xi)\big)^{1/2}$,
so that 
\begin{equation}
\label{OuterInt_CovProb}
\int_0^{\infty} \lambda(x) \, x^{\xi} \, f_{\kappa}(x) \, dx
= \left(\frac{2}{\kappa}\right)^{\xi/2} \frac{\Gamma(\nu/2)}{\Gamma(\kappa/2)}
\int_0^{\infty} a(y) \, f_{\nu}(y) \, dy,
\end{equation}
where $\nu = \kappa + \xi$ and  $a(y) = \lambda\big(y \big/ c(\kappa, \xi)\big)$ is a smooth bounded function of $y \ge 0$.

An important measure of the performance of a confidence interval is its coverage probability function. The assessment of the coverage probability functions of (a) the post-model-selection confidence intervals considered by \cite{KabailaGiri2009ANZJS}
and \cite{KabailaFarchione2012} and (b) the frequentist model averaged confidence intervals considered by \cite{KabailaWelshAbeysekera2016} is carried out by plotting the graphs of these functions. This requires the evaluation, for some given $\nu$, of an expression of the form (1) for, say, 200 different functions $a$. Each of these functions is smooth and bounded and the evaluation of $a(y)$ for any given $y$ is computationally epxensive.

\cite{AbeysekeraKabaila2017}, \cite{KabailaGiri2009JSPI},
\cite{KabailaGiri2013} and \cite{KabailaTissera2014} construct non-standard confidence regions with guaranteed coverage using the following computations. 
They numerically optimize a criterion related to the expected volume of a parametric family of non-standard confidence regions, subject to a coverage probabililty constraint. The coverage probabililty, for a particular true parameter value, of a member of this family is given by an expression of the form (1), for some given $\nu$.
The function $a$ is smooth and bounded and the
evaluation of $a(y)$ for any given $y$ is computationally expensive.
As this numerical constrained optimization proceeds, the evaluation of an expression of the form (1) 
needs to be carried out for thousands or even tens of thousands of different functions $a$, for the same given $\nu$.

\section{Remarks}

For the computation of $y_{\ell}$ and $d$ in Step 1 of our procedure,
we have evaluated $Q_{\nu}$ using the \texttt{R}  function \texttt{pchisq}. This evaluation of $Q_{\nu}$
is carried out using well-established methods for the evaluation of the incomplete gamma integral. These methods include the series expansion described by
\cite{Shea1988}, 
as well as a continued fraction expansion due to Gauss, which greatly simplifies for $\nu$ an even positive integer. 
As already noted, 
for the types of problems considered in the references co-authored
with Kabaila, the ``set-up cost'' of Step 1 is negligible.
However, if one really needed to reduce the computation time for Step 1 then one could do so by replacing the exact evaluation of the tail probabilities of the $\chi_{\nu}^2$ distribution 
by upper bounds (such as Chernoff bounds) on these probabilities and by simplifying the minimization and root finding steps needed to evaluate $y_{\ell}$ and $d$.

A reviewer has suggested, following Step 1 of our procedure, the application of Gauss-Legendre quadrature (instead of the application of the trapezoidal rule) to the evaluation of the truncated integral
\begin{equation}
\label{TruncatedIntegral}
\int_{y_{\ell}}^{y_u} a_{\nu, \alpha}\big(x(y) \big) \, \psi_{\nu}(y) \, dy. 
\end{equation}
This application is made in the usual way by first carrying out a straight line transformation of the 
interval $[y_{\ell}, y_u]$ to $[-1,1]$.
The resulting approximation errors are shown in Table 	\ref{table:GaussLegendreError}. The magnitudes of these approximation errors are all larger than the magnitudes of the corresponding approximation errors for the trapezoidal rule, with the same number of integrand evaluations, reported in Table 	\ref{table:TrapezoidError}. In other words, for the same number of evaluations of the integrand, the trapezoidal rule 
outperforms Gauss-Legendre quadrature applied to the evaluation of 
\eqref{TruncatedIntegral}, in terms of magnitude of approximation error. 
This result may be explained by the fact that the Gauss-Legendre quadrature nodes,
which lie in the interval $[-1, 1]$, cluster near the values $-1$ and 1, where the transformed integrand takes values very close to zero. This clustering also leads to Gauss-Legendre quadrature nodes not far from 0, where the transformed integrand differs most from zero, being more widely spaced than for the trapezoidal rule, with the same number
of evaluations of the integrand.

\FloatBarrier
%
\begin{table}[h] 
	\caption{The approximation error for Gauss Legendre quadrature applied to the evaluation of   \eqref{TruncatedIntegral},  for $\alpha = 0.10, 0.05$ and 0.02 and $\nu = 1, 2, 3, 4, 5, 10, 100$ and 1000. The number of nodes
		$m$  is $65$ for $\nu = 1$ and $33$ for $\nu = 2, 3, 4, 5, 10, 100$ and 1000.} 
	\label{table:GaussLegendreError} 
	\begin{tabular}{@{\extracolsep{0pt}} ccccccccc} 
		\hline 
		& $\nu=1$ & $\nu=2$ & $\nu=3$ & $\nu=4$ & $\nu=5$ \\ 
		\hline 
		$\alpha=0.10$ & $-2.84 \times 10^{-13}$ & $-2.02 \times 10^{-10}$ & $-1.26 \times 10^{-13}$ & $-2.47 \times 10^{-13}$ & $-6.75 \times 10^{-14}$ \\ 
		$\alpha=0.05$ & $-1.67 \times 10^{-10}$ & $-2.01 \times 10^{-8}$ & $-6.44 \times 10^{-11}$ & $8.88 \times 10^{-13}$ & $1.47 \times 10^{-12}$ \\ 
		$\alpha=0.02$ & $2.20 \times 10^{-7}$ & $-5.48 \times 10^{-7}$ & $-3.31 \times 10^{-9}$ & $1.70 \times 10^{-10}$ & $1.15 \times 10^{-12}$ \\ 
		\hline 
	\end{tabular}
	\begin{tabular}{@{\extracolsep{0.1pt}} ccccccccc} 
		\hline 
		& $\nu=10$ & $\nu=100$ & $\nu=1000$ \\ 
		\hline 
		$\alpha=0.10$ & $-3.80 \times 10^{-14}$ & $-9.33 \times 10^{-13}$ & $2.98 \times 10^{-13}$ \\ 
		$\alpha=0.05$ & $3.09 \times 10^{-14}$ & $-1.06 \times 10^{-12}$ & $3.74 \times 10^{-13}$ \\ 
		$\alpha=0.02$ & $6.85 \times 10^{-14}$ & $-1.13 \times 10^{-12}$ & $4.36 \times 10^{-13}$ \\ 
		\hline 
	\end{tabular}  
\end{table}
\FloatBarrier

\noindent An additional advantage of the trapezoidal rule is that, unlike Gauss-Legendre quadrature, it
leads to a nested sequence of quadrature rules that can be used for the estimation of the approximation error.

\section{Discussion}

In Section 4, for both Generalized Gauss Laguerre quadrature
and the Inverse cdf method, we present graphs whose features accurately predict their performance in terms of accuracy for a given number of evaluations of the function $a$. As noted in Section 5, the number of evaluations 
of the function $a$ is a reasonable measure of computational effort when the ``set-up costs'' are negligible, as in the situations considered in the references co-authored with Kabaila.

Our findings for the test scenario considered in Section 4 are as follows.
The Generalized Gauss Laguerre quadrature method performs worst for $\nu = 1$,
and has performance that improves with increasing $\nu$. It has the worst performance of the three methods for $\nu \in \{1,2,3,4\}$. 
The inverse cdf method, using Gauss Legendre quadrature, 
has the best performance of the three methods
$\nu = 1$ and $\alpha \in \{0.05, 0.02\}$, and has performance that decreases as $\nu$ increases through the values $2, 3, 4, 5, 6$ and 10.
The method described in Section 3 (application of the the transformation (2.6) of \citeauthor{Mori1988}, \citeyear{Mori1988}) has the best performance for $\nu = 1$ and $\alpha = 0.1$, $\nu \in \{1, 2, 3, 4, 5, 10\}$, has very close to the best performance for $\nu =100$
and has the best performance for $\nu = 1000$. 
For many of the situations considered in the references co-authored with Kabaila, the smallest possible value of $\nu$, in the evaluation of integrals of the form \eqref{FormofExpectedValue}, is 2. 

The procedures described in Section 3 use a nested sequence of quadrature rules, for the estimation of the approximation error, so that previous evaluations of the integrand are not wasted. This nested sequence can be implemented in a very simple computer program. This is an important advantage of this method over the other two methods. 

Taken together, the results presented in this paper show that 
the simple procedure described in subsection 3.1 
is a very suitable candidate for the computation of 
the coverage and expected volume properties of non-standard confidence regions considered in the references co-authored with Kabaila.  
The work presented in the paper is motivated by a need to compute
these properties. However, it is clear that the application of transformations, 
such as those put forward by \cite{Schwartz1969}, \cite{TakahasiMori1973}
and \cite{Mori1988}, followed by the application of the trapezoidal rule
will be useful in computing the expected values of functions of 
continuous 
random variables for a wide range of probability distributions of these random variables.

\bigskip

\noindent \textbf{Acknowledgement}

\smallskip

\noindent This work was supported by an Australian Government Research Training Program Scholarship. 


\begin{thebibliography}{31}
	\providecommand{\natexlab}[1]{#1}
	\providecommand{\url}[1]{{#1}}
	\providecommand{\urlprefix}{URL }
	\expandafter\ifx\csname urlstyle\endcsname\relax
	\providecommand{\doi}[1]{DOI~\discretionary{}{}{}#1}\else
	\providecommand{\doi}{DOI~\discretionary{}{}{}\begingroup
		\urlstyle{rm}\Url}\fi
	\providecommand{\eprint}[2][]{\url{#2}}
	
	\bibitem[{Abeysekera and Kabaila(2017)}]{AbeysekeraKabaila2017}
	Abeysekera W, Kabaila P (2017) Optimized recentered confidence spheres for the
	multivariate normal mean. Electronic Journal of Statistics 11:1935--7524
	
	\bibitem[{Avery and Soler(1988)}]{AverySoler1988}
	Avery C, Soler F (1988) Applications of transformations to numerical
	integration. The College Mathematics Journal 19:166--168
	
	\bibitem[{Chandrasekhar(1960)}]{Chandrasekhar1960}
	Chandrasekhar S (1960) Radiative transfer. Dover, New York
	
	\bibitem[{Davis and Rabinowitz(1984)}]{DavisRabinowitz1984}
	Davis PJ, Rabinowitz P (1984) Methods of Numerical Integration, 2nd edn.
	Academic Press, San Diego, CA
	
	\bibitem[{Dunnett(1989)}]{Dunnett}
	Dunnett CW (1989) Algorithm {AS} 251: Multivariate normal probability integrals
	with product correlation structure. Journal of the Royal Statistical Society,
	Series C (Applied Statistics) 38:564--579
	
	\bibitem[{Dunnett and Sobel(1955)}]{DunnettSobel}
	Dunnett CW, Sobel M (1955) Approximations to the probability integral and
	certain percentage points of a multivariate analogue of student's
	t-distribution. Biometrika 42:258--260
	
	\bibitem[{Farchione and Kabaila(2008)}]{FarchioneKabaila2008}
	Farchione D, Kabaila P (2008) Confidence intervals for the normal mean
	utilizing prior information. Statistics and Probability Letters 78:1094--1100
	
	\bibitem[{Genz and Bretz(2009)}]{GenzBretz}
	Genz A, Bretz F (2009) Computation of multivariate normal and t probabilities.
	Springer, London
	
	\bibitem[{Gupta and Panchapakesan(2002)}]{GuptaPanchapakesan2002}
	Gupta SS, Panchapakesan S (2002) Multiple decision procedures: theory and
	methodology of selecting and ranking populations. SIAM, Philadelphia
	
	\bibitem[{Hochberg and Tamhane(1987)}]{HochbergTamhane1987}
	Hochberg Y, Tamhane AC (1987) Multiple comparison procedures. Wiley, New York
	
	\bibitem[{Imhof(1963)}]{Imhof1963}
	Imhof J (1963) On the method for numerical integration of {C}lenshaw and
	{C}urtis. Numerische Mathematik 5:138--141
	
	\bibitem[{Kabaila(2018)}]{Kabaila2018}
	Kabaila P (2018) On the minimum coverage probability of model averaged tail
	area confidence intervals. Canadian Journal of Statistics 46:279--297
	
	\bibitem[{Kabaila and Farchione(2012)}]{KabailaFarchione2012}
	Kabaila P, Farchione D (2012) The minimum coverage of confidence intervals in
	regression after a preliminary {F} test. Journal of Statistical Planning and
	Inference 142:956--964
	
	\bibitem[{Kabaila and Giri(2009{\natexlab{a}})}]{KabailaGiri2009JSPI}
	Kabaila P, Giri K (2009{\natexlab{a}}) Confidence intervals in regression
	utilizing prior information. Journal of Statistical Planning and Inference
	139:3419--3429
	
	\bibitem[{Kabaila and Giri(2009{\natexlab{b}})}]{KabailaGiri2009ANZJS}
	Kabaila P, Giri K (2009{\natexlab{b}}) Upper bounds on the minimum coverage
	probability of confidence intervals in regression after model selection.
	Australian \& New Zealand Journal of Statistics 51:271--287
	
	\bibitem[{Kabaila and Giri(2013)}]{KabailaGiri2013}
	Kabaila P, Giri K (2013) Further properties of frequentist confidence intervals
	in regression that utilize uncertain prior information. Australian \& New
	Zealand Journal of Statistics 55:259--270
	
	\bibitem[{Kabaila and Tissera(2014)}]{KabailaTissera2014}
	Kabaila P, Tissera D (2014) Confidence intervals in regression that utilize
	uncertain prior information about a vector parameter. Australian \& New
	Zealand Journal of Statistics 56:371--383
	
	\bibitem[{Kabaila et~al.(2016)Kabaila, Welsh, and
		Abeysekera}]{KabailaWelshAbeysekera2016}
	Kabaila P, Welsh A, Abeysekera W (2016) Model-averaged confidence intervals.
	Scandinavian Journal of Statistics 43:35--48
	
	\bibitem[{Kabaila et~al.(2017)Kabaila, Welsh, and
		Mainzer}]{KabailaWelshMainzer2017}
	Kabaila P, Welsh A, Mainzer R (2017) The performance of model averaged tail
	area confidence intervals. Communications in Statistics - Theory and Methods
	46:10718--10732
	
	\bibitem[{Mi et~al.(2009)Mi, Miwa, and Hothorn}]{MiMiwaHothorn2009}
	Mi X, Miwa T, Hothorn T (2009) mvtnorm: new numerical algorithm for
	multivariate normal probabilities. R Journal 1:37--39
	
	\bibitem[{Miller(1981)}]{Miller1981}
	Miller RG (1981) Simultaneous statistical inference, 2nd ed. Springer, New York
	
	\bibitem[{Miwa et~al.(2003)Miwa, Hayter, and Kuriki}]{MiwaHayterKuriki2003}
	Miwa T, Hayter AJ, Kuriki S (2003) The evaluation of general non-centred
	orthonant probabilities. Journal of the Royal Statistical Society, Series B
	65:223--234
	
	\bibitem[{Mori(1985)}]{Mori1985}
	Mori M (1985) Quadrature formulas obtained by variable transformation and the
	de-rule. Journal of Computational and Applied Mathematics 12 \& 13:119--130
	
	\bibitem[{Mori(1988)}]{Mori1988}
	Mori M (1988) The double exponential formula for numerical integration over the
	half infinite interval. In: Agarwal R, Chow Y, Wilson S (eds) Numerical
	Mathematics (Singapore 1988), Birkhauser, Basel, pp 367--379
	
	\bibitem[{Mori and Sugihara(2001)}]{MoriSugihara2001}
	Mori M, Sugihara M (2001) The double-exponential transformation in numerical
	analysis. Journal of Computational and Applied Mathematics 127:287--296
	
	\bibitem[{Papoulis(1962)}]{Papoulis1962}
	Papoulis A (1962) The {F}ourier integral and its applications. McGraw-Hill, New
	York
	
	\bibitem[{Sag and Szekeres(1964)}]{SagSzekeres1964}
	Sag TW, Szekeres G (1964) Numerical evaluation of high-dimensional integrals.
	Mathematics of Computation 18:245--253
	
	\bibitem[{Schwartz(1969)}]{Schwartz1969}
	Schwartz C (1969) Numerical integration of analytic functions. Journal of
	Computational Physics 4:19--29
	
	\bibitem[{Shea(1988)}]{Shea1988}
	Shea B (1988) Algorithm {AS} 239: Chi-squared and incomplete gamma integral.
	Journal of the Royal Statistical Society, Series C 37:466--473
	
	\bibitem[{Takahasi and Mori(1973)}]{TakahasiMori1973}
	Takahasi H, Mori M (1973) Quadrature formulas obtained by variable
	transformation. Numerische Mathematik 21:206--219
	
	\bibitem[{Trefethen and Weideman(2014)}]{TrefethenWeidemanSIAMReview2014}
	Trefethen LN, Weideman JAC (2014) The exponentially convergent trapezoidal
	rule. SIAM Review 56:385--458
	
\end{thebibliography}

\end{document}